\documentclass[11pt,a4paper]{article}
\usepackage{authblk}
\usepackage{lmodern}
\usepackage{jstyle}
\usepackage{amsmath}
\usepackage{mathtools}
\usepackage{tikz}
\usepackage{empheq}
\usepackage{enumitem}
\usepackage{graphics}
\usepackage{wrapfig}
\usepackage{caption}
\usepackage{natbib}
\usepackage{xcolor}
\usepackage{framed}
\usepackage{array}
\definecolor{shadecolor}{gray}{0.925}

\usepackage[cbgreek]{textgreek}

\def\sideremark#1{\ifvmode\leavevmode\fi\vadjust{\vbox to0pt{\vss
 \hbox to 0pt{\hskip\hsize\hskip1em
 \vbox{\hsize3cm\tiny\raggedright\pretolerance10000
 \noindent #1\hfill}\hss}\vbox to8pt{\vfil}\vss}}}%

\newcommand{\bi}{\begin{itemize}}
\newcommand{\ei}{\end{itemize}}
\newcommand{\bea}{\begin{align}}
\newcommand{\eea}{\end{align}}
\newcommand{\be}{\begin{equation}}
\newcommand{\ee}{\end{equation}}

\makeatletter
\renewcommand*\env@matrix[1][\arraystretch]{%
  \edef\arraystretch{#1}%
  \hskip -\arraycolsep
  \let\@ifnextchar\new@ifnextchar
  \array{*\c@MaxMatrixCols c}}
\makeatother

\author{\, Charlotte SLEIGHT}
\author{\, Massimo TARONNA}

\affiliation{Dipartimento di Fisica ``Ettore Pancini'', Universit\`a degli Studi di Napoli Federico II, \\Monte S. Angelo, Via Cintia, 80126 Napoli, Italy}

\affiliation{INFN, Sezione di Napoli, Monte S. Angelo, Via Cintia, 80126 Napoli, Italy}

\emailAdd{charlotte.sleight@na.infn.it, massimo.taronna@unina.it}

\title{\centering \Huge (Non-)Conserved Currents\\ and Cosmological Correlators}

\abstract{We study the fate of global symmetries at the late-time boundary of de Sitter space. In anti-de Sitter space, bulk gauge symmetries generally correspond to conserved global currents on the boundary. We show that in de Sitter space such currents tend to acquire anomalous dimensions due to multiplet recombination with composite operators, which is a consequence of the shadow structure of the boundary operator spectrum. As a result, global symmetries are generically (weakly) broken. This mechanism is transparent in the EAdS reformulation \cite{Sleight:2020obc,Sleight:2021plv} of dS late-time correlators, where Dirichlet modes mix with composites and acquire small masses, while Neumann modes remain protected by gauge invariance. We demonstrate this mechanism explicitly in scalar QED, scalar minimally coupled to Gravity, Yang–Mills theory, and pure Einstein gravity, extracting the corresponding anomalous dimensions. We argue that it extends to higher-spin and partially massless fields.}

\begin{document}

\begin{flushright}    
\texttt{}
\end{flushright}

\maketitle

\section{Introduction}\label{sec::Intro}

In the anti-de Sitter (AdS)/Conformal Field Theory (CFT) correspondence \cite{Maldacena:1997re,Gubser:1998bc,Witten:1998qj}, bulk gauge symmetries manifest as global symmetries on the boundary \cite{Harlow:2018tng}. We show that at the late-time boundary of de Sitter (dS) space this paradigm is modified: boundary radiation fluctuations tend to spontaneously break all global symmetries, and the associated currents acquire anomalous dimensions through recombination with marginal double-trace operators.

\vskip 4pt
 Unlike in AdS, where the two asymptotic behaviors of a bulk field are interpreted as source and VEV, in de Sitter space late-time correlators naturally retain both, leading to boundary operators with complementary dimensions that reduce to shadow pairs in the free theory. At late times $\eta \to 0$, a field $\varphi_s$ of spin-$s$ in de Sitter space approaches the form:\footnote{See section \ref{subsec::notation and conventions} for notations and conventions.}
\begin{equation}\label{bdrybehaviour}
    \varphi_{s}\left(\eta \to 0,{\bf x}\right) = \left(-\eta\right)^{\Delta_+-s}{\cal O}_{\Delta_+,s}\left({\bf x}\right)+\left(-\eta\right)^{\Delta_--s}{\cal O}_{\Delta_-,s}\left({\bf x}\right),
\end{equation}
where the boundary operators ${\cal O}_{\Delta_\pm,s}({\bf x})$ are spin-$s$ conformal primaries with scaling dimensions $\Delta_\pm$.  In the free theory these satisfy the shadow relation $\Delta_++\Delta_-=d$ and are related to the particle mass via
\begin{equation}
    m^2 = \Delta_+\Delta_-+s.
\end{equation}
The contrast with AdS is that there the choice of Dirichlet or Neumann boundary conditions singles out one of the two fall-offs in \eqref{bdrybehaviour},\footnote{The same holds in the wavefunction approach \cite{Maldacena:2002vr}, where the dS/CFT dictionary identifies late-time wavefunctions with generating functionals of the would-be dual CFT, and one specifies a boundary condition for the bulk fields at future infinity. This is related by analytic continuation to the partition function in a Euclidean AdS background. In this paper we work with dS late-time correlators, which are the physical expectation values that can be obtained by applying the Born rule to the cosmological wavefunction.} whereas for dS late-time correlators the choice of initial state in the past fixes the dynamics and both late-time fall-offs contribute. 

\vskip 4pt
This shadow pairing plays a crucial role for gauge bosons, gravitons, and gauge fields more generally, as their boundary limits control the existence and fate of global symmetries. For a gauge boson $A_i$ and graviton $h_{ij}$, in the free theory we have\footnote{In this paper we work in the temporal gauge, which corresponds to $A_0 = 0$ and $h_{0 \nu}=0$.} 
\begin{align}\label{Agfree}
    A_i(\eta \to 0, {\bf x})&= {\tilde a}_i({\bf x})+\left(-\eta\right)^{d-2}J_i({\bf x}),\\
    h_{ij}(\eta \to 0, {\bf x})&= \eta^{-2}{\tilde h}_{ij}({\bf x})+\left(-\eta\right)^{d-2}T_{ij}({\bf x}).
\end{align}
The fields ${\tilde a}_i({\bf x})$ and ${\tilde h}_{ij}({\bf x})$ correspond, respectively, to a boundary gauge boson and graviton with scaling dimensions $\Delta_a=1$ and $\Delta_h=0$. $J_i({\bf x})$ and $T_{ij}({\bf x})$ have scaling dimensions $\Delta_J=d-1$ and $\Delta_T=d$ respectively and define short multiplets of the boundary conformal group, with $J_i$ associated to a spin-1 current and $T_{ij}$ the stress tensor of the boundary CFT. In AdS, the boundary fields ${\tilde a}_i$ and ${\tilde h}_{ij}$ would correspond to non-unitary operators, since their scaling dimensions lie below the unitarity bound \cite{Metsaev:1995re}. In de Sitter space, by contrast, both types of boundary operators are unitary representations of the isometry group \cite{Dobrev:1977qv}, and quantum fluctuations will mix them in a way that modifies the scaling of $J_i$ and $T_{ij}$.

\vskip 4pt
While the current $J_i({\bf x})$ and stress tensor $T_{ij}({\bf x})$ are classically conserved, the shadow structure of boundary CFT results in them acquiring anomalous dimensions $\gamma_J$ and $\gamma_T$ due to quantum fluctuations:
\begin{align}\label{anomdim}
    A_i(\eta \to 0, {\bf x})&= {\tilde a}_i({\bf x})+\left(-\eta\right)^{d-2+\gamma_J}J_i({\bf x}),\\
    h_{ij}(\eta \to 0, {\bf x})&= \eta^{-2}{\tilde h}_{ij}({\bf x})+\left(-\eta\right)^{d-2+\gamma_T}T_{ij}({\bf x}).
\end{align}
This is owing to the presence of ``double trace" operators in the free theory spectrum composed of shadow operators ${\cal O}_{\Delta_+}$ and ${\cal O}_{\Delta_-}$ associated \eqref{bdrybehaviour} to each field $\varphi$ in the bulk.\footnote{Note that the operators ${\cal O}_{\Delta_+}$ may themselves be the current $J_i$ or stress tensor $T_{ij}$, in which case ${\cal O}_{\Delta_-}$ corresponds to the boundary gauge boson ${\tilde a}_i$ or graviton ${\tilde h}_{ij}$.} These take the schematic form
\begin{equation}\label{DT}
    \left[{\cal O}_{\Delta_+} {\cal O}_{\Delta_-}\right]_{n,i_1 \ldots i_\ell} \sim {\cal O}_{\Delta_+} \partial_{i_1} \ldots  \partial_{i_\ell} (\partial^2)^n {\cal O}_{\Delta_-} + \ldots,
\end{equation}
with scaling dimensions
\begin{equation}\label{DTdim}
  \Delta_{n,\ell} =  d+2n+\ell.
\end{equation}
Given the scaling dimensions of $\Delta_J = d-1$ and $\Delta_T = d$ of the current and stress tensor, this suggests the possibility of a mixing between their divergence and the double-trace operators \eqref{DT}, resulting in the multiplet recombination:
\begin{subequations}\label{MR}
 \begin{align}
    \partial^i J_i &\sim g \left[{\cal O}_{\Delta_+} {\cal O}_{\Delta_-}\right]_{0} + O(g^2),\\
     \partial^i T_{ij} & \sim g \left[{\cal O}_{\Delta_+} {\cal O}_{\Delta_-}\right]_{0,j} + O(g^2),
\end{align}   
\end{subequations}
where $g$ is the bulk coupling constant. By contrast, the dimensions $\Delta_a=1$ and $\Delta_h=0$ of the boundary gauge boson ${\tilde a}_i$ and graviton ${\tilde h}_{ij}$ remain protected by gauge symmetry. Note that this phenomenon is \emph{universal} for theories of gauge bosons and gravitons due to minimal coupling of the bulk field $\varphi$ associated to $\mathcal{O}_{\Delta_\pm}$.

\begin{figure}[t]
    \centering
    \includegraphics[width=0.7\textwidth]{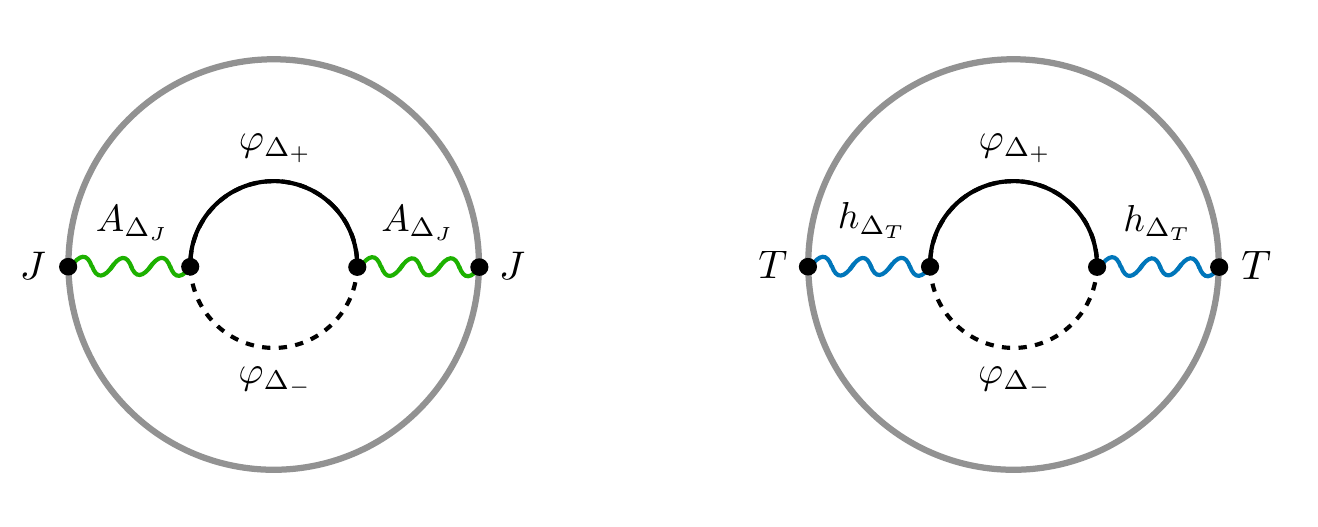}
    \caption{Loop correction to gauge boson $A_{\Delta_J}$ and graviton $h_{\Delta_J}$ mass in EAdS induced by a the bound state of fields $\varphi_{\Delta_\pm}$.}
    \label{fig::loopcorrections}
\end{figure}

\vskip 4pt
In this paper we verify the multiplet recombination \eqref{MR} explicitly for scalar QED, scalar minimally coupled to gravity, Einstein Gravity and Pure Yang-Mills theory in dS$_{d+1}$ with the standard Bunch-Davies initial conditions \cite{Bunch:1978yq}. This is demonstrated by evaluating the divergence of the corresponding three-point functions $\langle J_{i_1 \ldots i_s} {\cal O}_{\Delta_+} {\cal O}_{\Delta_-}\rangle $ perturbatively in the algebra of local operators, showing that conservation is violated by terms proportional to (derivatives of) the product of $\mathcal{O}_{\Delta_\pm}$ two-point functions:
  \begin{align} \label{div3pt}
    \left\langle \partial^i J_i \mathcal{O}_{\Delta_+}\mathcal{O}_{\Delta_-}\right\rangle&\sim g\left\langle\mathcal{O}_{\Delta_+}\mathcal{O}_{\Delta_+}\right\rangle\left\langle\mathcal{O}_{\Delta_-}\mathcal{O}_{\Delta_-}\right\rangle\,,\\ \nonumber
    \left\langle  \partial^i T_{ij} \mathcal{O}_{\Delta_+}\mathcal{O}_{\Delta_-}\right\rangle&\sim g\,\partial_j \left\langle\mathcal{O}_{\Delta_+}\mathcal{O}_{\Delta_+}\right\rangle\left\langle\mathcal{O}_{\Delta_-}\mathcal{O}_{\Delta_-}\right\rangle + \ldots\,+ O(g^2),
\end{align}
 where $\ldots$ denotes similar terms with derivative $\partial_j$.

\vskip 4pt
It is instructive to analyse this mechanism from the perspective of the perturbative reformulation \cite{Sleight:2020obc,Sleight:2021plv} of late-time dS correlators in terms of correlators on the boundary of Euclidean AdS (EAdS) space. A dedicated treatment for theories of gauge bosons and gravitons was recently given in \cite{MdAbhishek:2025dhx}. In this reformulation, each dS field $\varphi$ is replaced by a pair of EAdS fields $\varphi_{\Delta_\pm}$ corresponding to the two fall-offs \eqref{bdrybehaviour}. For gauge bosons and gravitons \eqref{Agfree}, this amounts to a pair of EAdS gauge bosons or gravitons associated to Dirichlet ($\Delta_+=\Delta_J$/$\Delta_T$) and Neumann ($\Delta_-=\Delta_a$/$\Delta_h$) boundary conditions. From this perspective, the non-conservation \eqref{MR} of the boundary current or stress tensor is naturally interpreted as a Higgs-like mechanism (see e.g. \cite{Porrati:2001db,Porrati:2003sa,Karch:2023wui}) induced by quantum corrections: loops of shadow fields $\varphi_{\Delta_\pm}$ generate a finite mass renormalisation for the Dirichlet gauge boson $A_{\Delta_J}$ or graviton $h_{\Delta_T}$ via mixing with the two-particle states \eqref{DT} (see figure \ref{fig::loopcorrections}). The Neumann modes, by contrast, remain massless due to gauge symmetry. This is the same mechanism that operates in AdS higher-spin theories dual to the critical $O(N)$ model, where boundary conditions explicitly break the higher-spin symmetry and generate small bulk masses for higher-spin fields \cite{Girardello:2002pp,Manvelyan:2008ks,Giombi:2011ya}. Related effects have recently been observed for partially massless spin-2 fields in conformal gravity \cite{Baumann:2025tkm}.

\begin{figure}[t]
    \centering
    \includegraphics[width=1\textwidth]{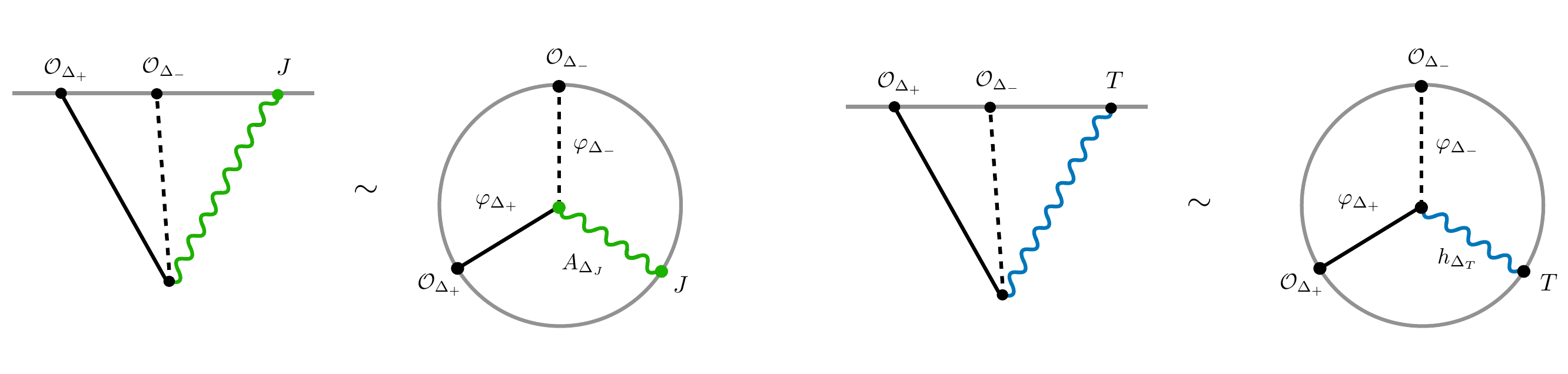}
    \caption{Non-conserved three-point functions of $J_i$ or $T_{ij}$ with shadow operators ${\cal O}_{\Delta_\pm}$ on the future boundary of de Sitter space can be recast as EAdS Witten diagrams involving gauge bosons $A_{\Delta_J}$ or gravitons $h_{\Delta_T}$ and the fields $\varphi_{\Delta_\pm}$.}
    \label{fig::3ptcont}
\end{figure}

\vskip 4pt
The EAdS reformulation not only clarifies the interpretation of the mechanism; it provides a practical tool for explicit calculations. While the standard approach to dS correlators employs the in-in (Schwinger–Keldysh) formalism \cite{Maldacena:2002vr,Bernardeau:2003nx,Weinberg:2005vy}, it is often convenient to recast them in terms of EAdS correlators. In particular, contact contributions to late-time dS correlators can be expressed directly in terms of corresponding contact Witten diagrams in EAdS \cite{Sleight:2020obc,Sleight:2021plv}:
\begin{multline}\label{sin contact}
    \langle {\cal O}_{\Delta_1, s_1} \ldots {\cal O}_{\Delta_n, s_n} \rangle_{\text{dS}\,\text{contact}} = \left(\prod^n_{i=1} c^{\text{dS-AdS}}_{\Delta_i} \right) 2\sin\left(\left(-d+N+\sum_i (\Delta_i- s_i)\right)\frac{\pi}{2}\right)\\
   \times \langle {\cal O}_{\Delta_1, s_1}  \ldots {\cal O}_{\Delta_n, s_n} \rangle_{\text{EAdS}\,\text{contact}},
\end{multline}
where $N$ is an integer that depends on the structure of the vertex, the coefficients $c^{\text{dS-AdS}}_{\Delta}$ account for the change in two point normalisation and the sinusoidal factor combines the contributions from each branch of the in-in contour. We will not review the details of this reformulation here and instead refer the reader to \cite{MdAbhishek:2025dhx}, which contains a dedicated treatment for gauge bosons and gravitons. In this language, non-conservation \eqref{MR} can be analysed through the corresponding contact Witten diagram in EAdS (see figure \ref{fig::3ptcont}), as long as the sinusoidal prefactor does not vanish.\footnote{A vanishing sinusoidal factor would instead indicate an exact cancellation between contributions from each branch of the in-in (Schwinger-Keldysh) contour.} One can express the anomalous dimensions $\gamma_{T/J}$ in terms of their EAdS counterparts generated by the loop processes in figure \ref{fig::loopcorrections} (see equation (5.25) of \cite{Sleight:2021plv}):
\begin{align}\label{dSEAdSanom}
        \gamma_{J/T} &= 4\frac{c^{\text{dS-AdS}}_{\Delta_+}c^{\text{dS-AdS}}_{\Delta_-}}{c^{\text{dS-AdS}}_{\Delta_{J/T}}}  \gamma^{\text{EAdS}}_{J/T}.
\end{align}
The corresponding correction to the mass of the bulk gauge boson/graviton in EAdS with Dirichlet boundary condition then follows from the standard relation:
\begin{equation}\label{massform}
    m^2 R^2_{\text{AdS}} = \left(\Delta+\gamma\right)\left(\Delta+\gamma-d\right) \sim \Delta\left(\Delta-d\right)+\gamma\left(2\Delta-d\right)+O(g^2).
\end{equation}

Although the focus of this work is on theories of gauge bosons and gravitons, this symmetry breaking mechanism extends to other types of gauge fields in dS$_{d+1}$, including higher-spin and partially-massless gauge fields. For example, for a massless spin-$s$ gauge field in dS$_{d+1}$, in the free theory we have
\begin{equation}
    \varphi_{i_1 \ldots i_s}\left(\eta\to0,{\bf x}\right) = \left(-\eta\right)^{2-2s}{\tilde \varphi}_{i_1 \ldots i_s}+\left(-\eta\right)^{d-2}J_{i_1 \ldots i_s},
\end{equation}
where the boundary current $J_{i_1 \ldots i_s}$ and boundary massless gauge field ${\tilde \varphi}_{i_1 \ldots i_s}$ have scaling dimensions:
\begin{equation}\label{mlspinsc}
    \Delta_{J_s} = s+d-2, \qquad \Delta_{{\tilde \varphi}_s} = 2-s.
\end{equation}
At the interacting level, higher-spin gauge theories in (A)dS consist of an infinite tower of massless gauge fields of all (even) integer spins \cite{Fradkin:1986qy}. This gives rise to a more general mixing of the form:
\begin{equation}
    \partial^{i_s} J_{i_1 \ldots i_s} \sim g \left[J_{s_1} {\tilde \varphi}_{s_2}\right]_{\frac{1}{2}(s_1-s_2),\,i_1 \ldots i_{s-1}}+ O(g^2), \qquad s_1 \geq s_2,
\end{equation}
which is induced the cubic interaction of the spin-$s$ field with other massless gauge fields of spins $s_1$ and $s_2$ in the higher spin multiplet. A similar discussion can also be made for partially-massless gauge fields, though it is more involved due to more complicated constraints from partial gauge symmetry on their possible non-trivial cubic interactions \cite{Joung:2012hz}. We therefore leave the discussion of the partially-massless case in appendix \ref{subsec::PM}. The complete classification of the cubic interactions of massless and partially massless totally symmetric fields in (A)dS$_{d+1}$ can be found in \cite{Joung:2011ww,Joung:2012rv,Joung:2012hz,Joung:2013nma}.

\vskip 4pt
It is instructive to place this phenomenon in the broader context of spontaneous symmetry breaking (SSB). In flat space, the only available mechanism is the standard Higgs effect: a scalar acquires a nonzero expectation value $\langle \Phi\rangle \neq 0$, which induces an explicit mass term in the effective action. AdS space also admits this mechanism, but holography provides a dual description in terms of multiplet recombination,
\begin{align}
\partial \cdot J \sim \mathcal{O}_\Phi ,
\end{align}
where the scalar vev corresponds to a single-trace marginal operator $\mathcal{O}_\Phi$, dual to the bulk Goldstone mode, that recombines with the current. In addition, AdS allows for recombination through mixing with composite operators (multi-trace states), which is the natural AdS counterpart of the mechanism we uncover in de Sitter. The crucial difference is that in dS the standard Higgs mechanism seems to be absent: analytic continuation to the sphere, with its finite volume, appears to forbid any local operator from acquiring a vev. Thus while both AdS and dS admit multiplet recombination, in AdS it can occur through scalar vevs or mixing with composites, whereas in dS it appears that only the latter is possible. Flat space, by contrast, only realises the Higgs mechanism in its standard form. Interestingly, all these aspects can still be discussed using the same language of AdS/CFT.

\vskip 4pt
The rest of the paper is organised as follows: In section \ref{sec::explicit} we explicitly verify the breaking of global symmetries in scalar QED, Yang-Mills theory and Einstein Gravity by calculating the divergence \eqref{div3pt} of the three-point function of the boundary current $J_i$ and stress tensor $T_{ij}$ with shadow operators ${\cal O}_{\Delta_\pm}$ at linear order in the bulk coupling constant. This is carried out using the Mellin space representation \cite{Sleight:2019mgd,Sleight:2019hfp,Sleight:2021plv} of (EA)dS boundary correlators, which is reviewed in appendix \ref{app::Mellinspace}. In section \ref{subsec::ad} we extract the corresponding anomalous dimensions \eqref{anomdim} of the boundary currents. We conclude in section \ref{sec::discussion} with a discussion on conceptual subtleties, including gauge choice, locality, and gauge invariance. Notations and conventions are given in section \ref{subsec::notation and conventions}.

\newpage 

\subsection{Notation and conventions}
\label{subsec::notation and conventions}

We employ Poincaré coordinates for both Euclidean AdS$_{d+1}$ and de Sitter space dS$_{d+1}$:
\begin{align}\label{poincare}
{\rm d}s^2_{\text{EAdS}} = R_{\text{AdS}}^2 \frac{{\rm d}z^2 + {\rm d}{\bf x}^2}{z^2},
\qquad
{\rm d}s^2_{\text{dS}} = R_{\text{dS}}^2 \frac{-{\rm d}\eta^2 + {\rm d}{\bf x}^2}{\eta^2},
\end{align}
with the radial variables taking values $z \in [0,\infty)$ and $\eta \in (-\infty,0]$, the latter describing the expanding patch of dS. The conformal boundary is reached in the limits $z \to 0$ or $\eta \to 0$, respectively. Unless explicitly indicated, we will set the curvature radii $R_{\text{(A)dS}}$ to unity. Throughout, spacetime indices are denoted by Greek letters $\mu = 0,1,\ldots,d$, while spatial indices are indicated by Latin letters $i = 1,\ldots,d$.

\vskip 4pt
The spatial vector ${\bf x}$ parameterises the boundary directions. In these directions it is useful to work in Fourier space with spatial momenta ${\bf k}$, which makes manifest the translation symmetry. For a function $f({\bf x})$ and its Fourier transform ${\hat f}({\bf k})$ we have,
\begin{equation}\label{FT}
    f({\bf x}) = \int \frac{{\rm d}^d {\bf k}}{\left(2\pi\right)^d}\, e^{i {\bf x} \cdot {\bf k} } {\hat f}({\bf k}), \qquad {\hat f}({\bf k}) = \int {\rm d}^d {\bf x}\, e^{-i {\bf k} \cdot {\bf x} } f({\bf x}).
\end{equation}
We often denote the magnitude of ${\bf k}$ by $k=|${\bf k}$|$.

\vskip 4pt
In the bulk directions, where translation invariance is absent, it is natural to adopt a basis that diagonalises the dilatation operator. This can be implemented through the Mellin representation (see \cite{Sleight:2021plv} and references therein), in which the coordinates $z$ (or $\eta$ in dS) are exchanged for a Mellin variable $s$. For a given function $f(z)$ and its Mellin transform $\tilde f(s)$, one has:
\begin{equation}\label{MT}
f(z) = \int_{-i\infty}^{+i\infty} \frac{{\rm d}s}{2\pi i} 2 \tilde f(s) z^{-\left(2s-\tfrac{d}{2}\right)},
\qquad
\tilde f(s) = \int_0^\infty \frac{{\rm d}z}{z} f(z) z^{2s-\tfrac{d}{2}}.
\end{equation}
The contour of integration is selected such that it separates the poles of the $\Gamma$ functions. For convenience, we will sometimes use the shorthand notation
\begin{equation}
\Gamma(a\pm b) \equiv \Gamma(a+b)\Gamma(a-b).
\end{equation}

\vskip 4pt
It will often be useful to employ an index-free notation for boundary operators. Given a symmetric, spin-$s$ boundary tensor operator ${\cal O}_{i_1 \cdots i_s}\left({\bf k}\right)$, we introduce constant auxiliary vectors $\epsilon^i$ and write
\begin{equation}
    {\cal O}\left({\bf k};{\bm \epsilon}\right) = {\cal O}_{i_1 \cdots i_s}\left({\bf k}\right) \epsilon^{i_1} \ldots \epsilon^{i_s}.
\end{equation}
The traceless part of the original tensor can be extracted by acting
with the Thomas differential operator \cite{aeb47b77-ed7f-3f96-bcdb-fa5606d7bf1d}:
\begin{equation}\label{ThomasD}
    D^i_{{\bm \epsilon}} = \left(\frac{d}{2}-1 + {\bm \epsilon} \cdot \partial_{{\bm \epsilon}}\right)\partial^i_{{\bm \epsilon}}-\frac{1}{2} {\bm \epsilon}^i \partial^2_{{\bm \epsilon}}.
\end{equation}

\newpage

\section{Non-conservation of currents and the stress tensor}
\label{sec::explicit}

In this section we verify the breaking of global symmetries explicitly in scalar QED (section \ref{subsec::scalarqed}), a scalar field minimally coupled to gravity (section \ref{subsec::scmingrav}), pure Yang-Mills theory (section \ref{subsec::YM}) and pure Einstein Gravity (section \ref{subsec::EG}), at leading order in the coupling constant. 

\vskip 4pt
This is demonstrated by the considering the contact diagram contributions to the boundary three-point function of the currents with a shadow pair of operators ${\cal O}_{\Delta_\pm}$ in the boundary operator spectrum. To this end it is useful to employ the Mellin space representation \cite{Sleight:2021plv} of boundary correlators, where the divergence of currents can be extracted from the residues of a finite number of poles \cite{Sleight:2021iix}. The relevant aspects of the Mellin representation are reviewed in appendix \ref{app::Mellinspace}.

\subsection{Scalar QED}
\label{subsec::scalarqed}

We begin with scalar QED in dS$_{d+1}$, which has the following Lagrangian:
\begin{align}
    {\cal L}=&-\frac{1}{4}F^{\mu \nu}F_{\mu\nu}-\left(D^\mu\varphi\right)^\dagger\left(D_\mu\varphi\right)-m^2\varphi^\dagger\varphi,
\end{align}
where $D_\mu=\nabla_\mu+ieA_\mu$ and $m^2 = \Delta_+ \Delta_-$. In the temporal gauge $A_0=0$ the cubic and quartic vertices read: 
\begin{align}\label{cubicqed}
    &V_{A\varphi\varphi^\dagger}=ie \left(-\eta\right)^2\delta^{ij}A_i\left(\varphi^\dagger \partial_j \varphi - \varphi \partial_j \varphi^\dagger\right),\\
    &V_{AA\varphi\varphi^\dagger}=-e^2\left(-\eta\right)^2\delta^{ij}A_i A_j\varphi^\dagger \varphi.
\end{align}
At late times $\eta \to 0$ we have the operators: 
\begin{align}
    \varphi\left(\eta \to 0, {\bf x}\right) &= \left(-\eta\right)^{\Delta_+} {\cal O}_{\Delta_+}\left({\bf x}\right)+ \left(-\eta\right)^{\Delta_-} {\cal O}_{\Delta_-}\left({\bf x}\right),\\
    \varphi^\dagger\left(\eta \to 0, {\bf x}\right) &= \left(-\eta\right)^{\Delta_+} {\cal O}^\dagger_{\Delta_+}\left({\bf x}\right)+ \left(-\eta\right)^{\Delta_-} {\cal O}^\dagger_{\Delta_-}\left({\bf x}\right),\\
    A\left(\eta \to 0, {\bf x};{\bm \epsilon }\right) &= {\tilde a}\left({\bf x};{\bm \epsilon }\right)+ \left(-\eta\right)^{d-2} J\left({\bf x};{\bm \epsilon }\right),
\end{align}
where ${\cal O}_{\Delta_\pm}\left({\bf x}\right)$ and ${\cal O}^\dagger_{\Delta_\pm}\left({\bf x}\right)$ are Hermitian conjugate scalar operators with scaling dimensions $\Delta_\pm = \frac{d}{2}\pm i\mu$. The field ${\tilde a}\left({\bf x};{\bm \epsilon }\right)$ is the boundary gauge boson and $J\left({\bf x};{\bm \epsilon }\right)$ the boundary $U(1)$ current, which is classically conserved with scaling dimension $\Delta_J = d-1$.

\vskip 4pt
To study the quantum corrections to $J_i\left(\bf x\right)$, we consider its three-point function with ${\cal O}_{\Delta_\pm}\left({\bf x}\right)$ and ${\cal O}^\dagger_{\Delta_\pm}\left({\bf x}\right)$.\footnote{The three-point function of $J_i\left(\bf x\right)$ with two insertions of ${\cal O}_{\Delta_\pm}\left({\bf x}\right)$ or two insertions of ${\cal O}^\dagger_{\Delta_\pm}\left({\bf x}\right)$ is vanishing.} The leading contribution in perturbation theory is the contact diagram contribution generated by the cubic vertex \eqref{cubicqed}. This can be expressed in terms of the corresponding contact Witten diagram in EAdS via \cite{MdAbhishek:2025dhx}:
\begin{multline}
    \langle {\cal O}_{\Delta_1}\left({\bf x}_1\right){\cal O}^\dagger_{\Delta_2}\left({\bf x}_2\right)J\left({\bf x}_3;{\bm \epsilon_3 }\right)\rangle = 2 \sin\left(\left(\Delta_1+\Delta_2\right)\tfrac{\pi}{2} \right) c^{\text{dS-AdS}}_{\Delta_1}c^{\text{dS-AdS}}_{\Delta_2} c^{\text{dS-AdS}}_{\Delta_J} \\ \times \langle  {\cal O}_{\Delta_1}\left({\bf x}_1\right){\cal O}^\dagger_{\Delta_2}\left({\bf x}_2\right)J\left({\bf x}_3;{\bm \epsilon_3 }\right)\rangle_{\text{EAdS},\,\text{contact}} + O(e^2).
\end{multline}
To study conservation it is convient to work in Fourier space \eqref{FT}, where the scalar and gauge boson bulk-to-boundary propagators take the form \cite{Gubser:1998bc,Marotta:2024sce}:
\begin{align}\label{scbubo}
    K^{\text{AdS}}_{\Delta}\left(z;{\bf k}\right)&= \left(\frac{k}{2}\right)^{\Delta-\frac{d}{2}}\frac{z^{\frac{d}{2}}}{\Gamma(\Delta-\frac{d}{2}+1)}K_{\Delta-\frac{d}{2}}(kz),\\ \label{gbbubo}
     K^{\text{AdS}}_{\Delta,\,i;j}\left(z;{\bf k}\right)&=\pi_{ij}\frac{1}{\Gamma (\Delta-\frac{d}{2} +1)} \left(\frac{k}{2}\right)^{\Delta-\frac{d}{2}} z^{\frac{d}{2}-1} K_{\Delta-\frac{d}{2}}(k z)\,+\frac{k_i k_j}{\left( 2\Delta-d\right)k^2}z^{d-\Delta -1},
\end{align}
where $\pi_{ij}$ is the transverse projector and we take $\Delta=\Delta_J$ for insertions of $J_i({\bf x})$ and $\Delta=\Delta_a$ for insertions of ${\tilde a}_i({\bf x})$. Employing the Mellin representation \eqref{BesselKMellin} of the Bessel-$K$ function, one can establish the Mellin amplitude \eqref{MellinA} for the EAdS contact Witten diagram, which reads 
\begin{multline}\label{MA00J}
    {\cal A}^{\text{EAdS},\,\text{contact}}_{\Delta_1 \Delta_2 \Delta_J}\left(s_1,{\bf k}_1;s_2,{\bf k}_2;s_3,{\bf k}_3,{\bm \epsilon}_3\right) = ie\frac{\pi^{\frac{3}{2}}}{\Gamma \left(\frac{d}{2}\right) \Gamma (\Delta_1-\frac{d}{2}+1) \Gamma (\Delta_2-\frac{d}{2}+1)} \\ \times \lim_{z_0 \to 0} \left[i
    \left({\bm \epsilon}_3\cdot \textbf{k}_{12}+\frac{(s_2-s_1)}{s_3+\tfrac{d}4-\tfrac{3}{2}}\,{\bm \epsilon}_3\cdot \textbf{k}_3\right) \frac{z_0^{\frac{d+2}{2}-2s_1-2s_2-2s_3}}{\frac{d+2}{2}-2s_1-2s_2-2s_3}
         \right].
\end{multline}
It is straightforward to evaluate the divergence of the current $J_i$, which for the possible combinations of $\Delta_1 = \Delta_\pm$ and $\Delta_2 = \Delta_{\hat \pm}$ reads 
\begin{multline} \nonumber
\left({\bf k}_3 \cdot \partial_{{\bm \epsilon}_3}\right){\cal A}^{\text{EAdS},\,\text{contact}}_{\Delta_+ \Delta_+ \Delta_J}\left(s_1,{\bf k}_1;s_2,{\bf k}_2;s_3,{\bf k}_3,{\bm \epsilon}_3\right) = ie\frac{\pi^{\frac{3}{2}}}{\Gamma \left(\frac{d}{2}\right) \Gamma (i\mu+1) \Gamma (i\mu+1)} \\ \times  \lim_{z_0 \to 0}\left[i(d-2) (s_1-s_2) z_0^{\frac{d}{2}-1-2(s_1+s_2+s_3)}\right],
\end{multline}
\begin{multline}
    \left({\bf k}_3 \cdot \partial_{{\bm \epsilon}_3}\right){\cal A}^{\text{EAdS},\,\text{contact}}_{\Delta_- \Delta_- \Delta_J}\left(s_1,{\bf k}_1;s_2,{\bf k}_2;s_3,{\bf k}_3,{\bm \epsilon}_3\right)=ie\frac{\pi^{\frac{3}{2}}}{\Gamma \left(\frac{d}{2}\right) \Gamma (1-i\mu) \Gamma (1-i\mu)}\\ \times \lim_{z_0 \to 0} \left[i(d-2) (s_1-s_2) z_0^{\frac{d}{2}-1-2(s_1+s_2+s_3)}\right]\,, 
\end{multline} 
\begin{multline}
\left({\bf k}_3 \cdot \partial_{{\bm \epsilon}_3}\right){\cal A}^{\text{EAdS},\,\text{contact}}_{\Delta_+ \Delta_- \Delta_J}\left(s_1,{\bf k}_1;s_2,{\bf k}_2;s_3,{\bf k}_3,{\bm \epsilon}_3\right)=ie\frac{\pi^{\frac{3}{2}}}{\Gamma \left(\frac{d}{2}\right) \Gamma (i\mu+1) \Gamma (1-i\mu)}\\ \times \lim_{z_0 \to 0} \left[i\left((d-2) (s_1-s_2)+ i\mu\left(\tfrac{d}2-1- 2s_3\right)\right)z_0^{\frac{d}{2}-1-2(s_1+s_2+s_3)}\right],
\end{multline}
where the case $\Delta_1=\Delta_2$ was already given in \cite{Sleight:2021iix}. The inverse Mellin transform can be evaluated using Cauchy's theorem. Only a finite number of the poles \eqref{MBrep3pt} in the Mellin variables $s_i$ have a non-vanishing residue in the limit $z_0 \to 0$. These are at:
\begin{equation}
    s_1 = \pm \frac{i \mu}{2} , \qquad  s_2 = \mp \frac{i \mu}{2}, \qquad  s_3 =\frac{1}{2} \left(\frac{d}{2}-1\right). 
\end{equation}
For the three-point function of $J_i$ with operators $O_{\Delta_\pm}$ and $O^\dagger_{\Delta_\pm}$ with the \emph{same} scaling dimension, evaluating the residues of the above poles recovers the usual Ward-Takahashi identity for current conservation (see \cite{Sleight:2021iix} equation (2.50)):
\begin{multline}
   \left({\bf k}_3 \cdot \partial_{{\bm \epsilon}_3}\right) \langle {\cal O}_{\Delta_\pm}\left({\bf k}_1\right){\cal O}^\dagger_{\Delta_\pm}\left({\bf k}_2\right)J\left({\bf k}_3;{\bm \epsilon_3 }\right)\rangle= -\frac{1}{8}ie \sin \left( \pi \Delta_\pm \right)\csc\left(\tfrac{\pi d}{2} \right) \text{csch}\left(\pi  \mu\right)\\ \times \left[ \langle {\cal O}_{\Delta_\pm}\left({\bf k}_1\right){\cal O}^\dagger_{\Delta_\pm}\left(-{\bf k}_1\right) \rangle - \langle {\cal O}_{\Delta_\pm}\left({\bf k}_2\right){\cal O}^\dagger_{\Delta_\pm}\left(-{\bf k}_2\right) \rangle\right]+O(e^2)\,.
\end{multline}
The two-point function of ${\cal O}_{\Delta_\pm}$ and ${\cal O}^\dagger_{\Delta_\pm}$ is 
\begin{align}\label{2ptscqed}
    \langle {\cal O}_{\Delta_\pm}\left({\bf k}\right){\cal O}^\dagger_{\Delta_\pm}\left(-{\bf k}\right) \rangle=\frac{\Gamma(\mp i\mu)^2}{4\pi}\left(\frac{k}2\right)^{\pm 2i\mu}\,, \qquad \Delta_\pm = \frac{d}{2}\pm i\mu.
\end{align}
If the operators $O_{\Delta_\pm}$ and $O^\dagger_{\Delta_\pm}$ instead have \emph{shadow} scaling dimensions, one finds, up to contact terms: 
\begin{multline}\label{nc3ptscqed}
   \left({\bf k}_3 \cdot \partial_{{\bm \epsilon}_3}\right) \langle {\cal O}_{\Delta_+}\left({\bf k}_1\right){\cal O}^\dagger_{\Delta_-}\left({\bf k}_2\right)J\left({\bf k}_3;{\bm \epsilon_3 }\right)\rangle\\= -ie \frac{\mu}{2}\langle {\cal O}_{\Delta_+}\left({\bf k}_1\right){\cal O}^\dagger_{\Delta_+}\left(-{\bf k}_1\right) \rangle \langle {\cal O}_{\Delta_-}\left({\bf k}_2\right){\cal O}^\dagger_{\Delta_-}\left(-{\bf k}_2\right)\rangle +O(e^2)\,.
\end{multline}
The right hand side signals the non-conservation of $J_i$ induced by multiplet recombination 
\begin{equation}
     \partial\cdot J=\alpha_J [\mathcal{O}_{\Delta_+}\mathcal{O}^\dagger_{\Delta_-}]_{0}+\beta_J [\mathcal{O}^\dagger_{\Delta_+}\mathcal{O}_{\Delta_-}]_{0}+O(e^2),
\end{equation}
with double trace operators
\begin{equation}
    [\mathcal{O}_{\Delta_+}\mathcal{O}^\dagger_{\Delta_-}]_{0} = \mathcal{O}_{\Delta_+}\mathcal{O}^\dagger_{\Delta_-}, \qquad [\mathcal{O}^\dagger_{\Delta_+}\mathcal{O}_{\Delta_-}]_{0} = \mathcal{O}^\dagger_{\Delta_+}\mathcal{O}_{\Delta_-}.
\end{equation}
The coefficient $\alpha$ can be determined by plugging into the deformed Ward identity \eqref{nc3ptscqed}, which gives:
\begin{align}\label{alphaJqed}
    \alpha_J &= -e \frac{\mu}{2}.\\
    \beta_J &= +e \frac{\mu}{2}.
\end{align}

\subsection{Scalar field minimally coupled to gravity}
\label{subsec::scmingrav}

In this section we consider scalar field minimally coupled to gravity in dS$_{d+1}$, which has the following Lagrangian 
\begin{align}
\mathcal{L}=-\frac12(\nabla_\mu\phi)(\nabla^\mu\phi)-\frac{m^2}{2}\phi^2\,,
\end{align}
where $\nabla_\mu$ is the covariant derivative and $m^2=\Delta_+\Delta_-$. In the temporal gauge the cubic vertex in the weak field expansion in $h_{\mu \nu}$ around the dS background is: 
\begin{align}\label{mscgr3}
    {\cal V}_{\phi \phi h} &=-\kappa \left(-\eta\right)^4\,\delta^{i_1 j_1}\delta^{i_2 j_2}h_{j_1j_2} \left(\partial_{i_1}\phi \partial_{i_2} \phi-\frac1{2} \delta_{i_1 i_2}\eta^{-2}((\partial\phi)^2 +m^2\phi^2) \right).
\end{align}
On the late time boundary we have the operators: 
\begin{align}
    \varphi\left(\eta \to 0, {\bf x}\right) &= \left(-\eta\right)^{\Delta_+} {\cal O}_{\Delta_+}\left({\bf x}\right)+ \left(-\eta\right)^{\Delta_-} {\cal O}_{\Delta_-}\left({\bf x}\right),\\
    h\left(\eta \to 0, {\bf x};{\bm \epsilon }\right) &= \eta^{-2}{\tilde h}\left({\bf x};{\bm \epsilon }\right)+ \left(-\eta\right)^{d-2} T\left({\bf x};{\bm \epsilon }\right),
\end{align}
where ${\cal O}_{\Delta_\pm}\left({\bf x}\right)$ are scalar operators with scaling dimensions $\Delta_\pm=\frac{d}{2}\pm i\mu$, the field ${\tilde h}\left({\bf x};{\bm \epsilon }\right)$ is the boundary graviton and $T\left({\bf x};{\bm \epsilon }\right)$ the boundary stress tensor, which is classically conserved with scaling dimension $\Delta_T = d$.

\vskip 4pt
To study quantum corrections to $T_{ij}\left({\bf x}\right)$, we consider its three-point function with scalar operators ${\cal O}_{\Delta_+}\left({\bf x}\right)$ and ${\cal O}_{\Delta_-}\left({\bf x}\right)$. The contact diagram contribution generated by the cubic vertex \eqref{mscgr3} can be expressed in terms of the corresponding contact Witten diagram in EAdS via \cite{MdAbhishek:2025dhx}:
\begin{multline}
    \langle {\cal O}_{\Delta_1}\left({\bf k}_1\right){\cal O}_{\Delta_2}\left({\bf k}_2\right)T\left({\bf k}_3;{\bm \epsilon_3 }\right)\rangle = 2 \sin\left(\left(\Delta_1+\Delta_2\right)\tfrac{\pi}{2} \right) c^{\text{dS-AdS}}_{\Delta_1}c^{\text{dS-AdS}}_{\Delta_2} c^{\text{dS-AdS}}_{\Delta_T} \\ \times \langle  {\cal O}_{\Delta_1}\left({\bf k}_1\right){\cal O}_{\Delta_2}\left({\bf k}_2\right)T\left({\bf k}_3;{\bm \epsilon_3 }\right)\rangle_{\text{EAdS},\,\text{contact}} + O(\kappa^2),
\end{multline}
where $\Delta_{1,2}=\Delta_\pm$. To assemble the Witten diagram, in addition to the scalar bulk-to-boundary propagator \eqref{scbubo}, we also need the graviton bulk-to-boundary propagator which was recently given in terms of the Bessel-$K$ function in \cite{MdAbhishek:2025dhx}, equation (3.55). Employing the Mellin-Barnes representation \eqref{BesselKMellin} of the Bessel-$K$ function one can determine the Mellin amplitude \eqref{MellinA}: 
\begin{multline}\label{00TMA}
   {\cal A}^{\text{EAdS},\,\text{contact}}_{\Delta_+ \Delta_- \Delta_T}\left(s_1,{\bf k}_1;s_2,{\bf k}_2;s_3,{\bf k}_3,{\bm \epsilon}_3\right)=-\kappa\frac{\pi^{\frac{3}{2}}}{\Gamma \left(\frac{d}{2}+1\right) \Gamma (1+i\mu) \Gamma (1-i\mu)}\\ \times \lim_{z_0 \to 0}\left[\frac{1}2
    {\bm \epsilon}_3\cdot \textbf{k}_{12}\left({\bm \epsilon}_3\cdot \textbf{k}_{12}+\frac{(s_2-s_1)}{s_3+\tfrac{d}4-1}\,{\bm \epsilon}_3\cdot \textbf{k}_3\right)\right.\\ \left.-\frac{i \mu  {\bm \epsilon}_3\cdot\textbf{k}_12 {\bm \epsilon}_3\cdot\textbf{k}_3 \left(\frac{d+4}{2}-2 (s_1+s_2+s_3)\right)}{d (s_3+\frac{d}{4}-1)}+({\bm \epsilon}_3\cdot \textbf{k}_3)^2\Big(\ldots\Big)\right]z_0^{\tfrac{d+4}{2} - 2 (s_1 + s_2 + s_3)},
\end{multline}
where the $({\bm \epsilon}_3\cdot \textbf{k}_3)^2$ only contribute longitudinal terms to the divergence, which we omit in the following for ease of presentation. Taking the divergence of $T_{ij}$ then gives
\begin{multline}\label{anomWscgrav}
  \left({\bf k}_3 \cdot D_{{\bm \epsilon}_3}\right)  {\cal A}^{\text{EAdS},\,\text{contact}}_{\Delta_+ \Delta_- \Delta_T}\left(s_1,{\bf k}_1;s_2,{\bf k}_2;s_3,{\bf k}_3,{\bm \epsilon}_3\right)=\kappa\frac{\pi^{\frac{3}{2}}}{\Gamma \left(\frac{d}{2}\right) \Gamma (i\mu+1) \Gamma (1-i\mu)}{\bm \epsilon}_3\cdot {\bf k}_{12}\\ \times \lim_{z_0 \to 0} \left[\left(d (s_1-s_2)+ i\mu\left(d- 4s_3\right)\right)z_0^{\frac{d}{2}-2(s_1+s_2+s_3)}\right],
\end{multline}
where $D_{{\bm \epsilon}_3}$ is the Thomas derivative \eqref{ThomasD}. As before the inverse Mellin transform is evaluated by taking the residue of a finite number of poles in the Mellin representation \eqref{MellinA}, which in this case are at:
\begin{equation}
    s_1 = \pm \frac{i \mu}{2} , \qquad  s_2 = \mp \frac{i \mu}{2}, \qquad  s_3 =\frac{d}{4}.
\end{equation}
Up to longitudinal and contact terms this gives
\begin{multline}
   \left({\bf k}_3 \cdot D_{{\bm \epsilon}_3}\right) \langle {\cal O}_{\Delta_+}\left({\bf k}_1\right){\cal O}_{\Delta_-}\left({\bf k}_2\right)T\left({\bf k}_3;{\bm \epsilon_3 }\right)\rangle\\=-\kappa \frac{i\mu}{2}{\bm \epsilon}_3\cdot {\bf k}_{12}\langle {\cal O}_{\Delta_+}\left({\bf k}_1\right){\cal O}_{\Delta_+}\left(-{\bf k}_1\right) \rangle \langle {\cal O}_{\Delta_-}\left({\bf k}_2\right){\cal O}_{\Delta_-}\left(-{\bf k}_2\right)\rangle+O(\kappa^2).
\end{multline}
This signals non-conservation of $T_{ij}$ in the form: 
\begin{align}
    \partial^{i} T_{ij} =  \alpha_T \left[{\cal O}_{\Delta_+} {\cal O}_{\Delta_-}\right]_{0,j}+O(\kappa^2)\,,
\end{align}
arising from multiplet recombination with the spin-1 double-trace operator
\begin{equation}\label{DTscgrav}
    \left[{\cal O}_{\Delta_+} {\cal O}_{\Delta_-}\right]_{0,1}\left({\bf k}_1+{\bf k}_2;{\bm \epsilon}_3\right) = i{\bm \epsilon}_3\cdot {\bf k}_{12} {\cal O}_{\Delta_+}\left({\bf k}_1\right){\cal O}_{\Delta_-}\left({\bf k}_2\right),
\end{equation}
which we give up to longitudinal terms. The anomalous Ward identity \eqref{anomWscgrav} fixes $\alpha$:
\begin{equation}\label{scgravalpha}
    \alpha_T =- \kappa \frac{i\mu}{2}. 
\end{equation}

\subsection{Pure Yang-Mills}
\label{subsec::YM}

In this section we consider pure Yang–Mills theory with gauge group $SU(N)$. The corresponding Lagrangian takes the form
\begin{align}
    {\cal L}=&-\frac{1}{2}\text{tr}(F^{\mu\nu}F_{\mu\nu}),
\end{align}
where, 
\begin{subequations}
 \begin{align}
    D_\mu&:=\partial_\mu-i\mathrm{g}A_\mu,\\
    F_{\mu\nu}&=t^{\mathrm{a}}F^\mathrm{a}_{\mu\nu}:=\frac{i}{\mathrm{g}}[D_\mu,D_\nu]=\partial_\mu A_\nu-\partial_\nu A_\mu -i\mathrm{g}[A_\mu,A_\nu],
\end{align}   
\end{subequations}
with generators $t^\mathrm{a}\in SU(N)$.\footnote{These are normalised such that $\text{tr}(t^\mathrm{a}t^\mathrm{b})=\frac{\delta^{\mathrm{{ab}}}}{2}$ and $[t^\mathrm{a},t^\mathrm{b}]=if^{\mathrm{{abc}}}t^\mathrm{c}$.} Working in the temporal gauge $A_0 = 0$, the interaction terms give rise to the following three–gluon and four–gluon vertices:
\begin{subequations}\label{YM vertices}
 \begin{align}\label{YM cubic}
    V_{AAA}\left(\eta\right)=& -\mathrm{g}f^{\mathrm{{abc}}}\left(-\eta\right)^4\delta^{ik}\delta^{jl}A^{\mathrm{a}}_k A^{\mathrm{b}}_l(\partial_i A^{\mathrm{c}}_j),\\ \label{YM quartic}
    V_{AAAA}\left(\eta\right)=& -\frac{\mathrm{g}^2}{4}f^{\mathrm{{abe}}}f^{\mathrm{cde}}\left(-\eta\right)^4\delta^{ik}\delta^{jl}A^{\mathrm{a}}_i A^{\mathrm{b}}_j A^{\mathrm{c}}_k A^{\mathrm{d}}_l.
\end{align}   
\end{subequations}
On the late-time boundary we have the operators: 
\begin{align}
    A\left(\eta \to 0, {\bf x};{\bm \epsilon }\right) &= {\tilde a}\left({\bf x};{\bm \epsilon }\right)+ \left(-\eta\right)^{d-2} J\left({\bf x};{\bm \epsilon }\right),
\end{align}
where ${\tilde a}\left({\bf x};{\bm \epsilon }\right)$ is the boundary gauge boson and $J\left({\bf x};{\bm \epsilon }\right)$ the boundary current which is classically conserved with scaling dimension $\Delta_J = d-1$.

\vskip 4pt
To study quantum corrections to $J\left({\bf x};{\bm \epsilon }\right)$ we consider its three-point function with itself and a boundary gauge field ${\tilde a}\left({\bf x};{\bm \epsilon }\right)$. The contact diagram contribution to the latter, generated by the cubic vertex \eqref{YM cubic}, can be expressed in terms of the corresponding contact Witten diagram in EAdS via \cite{MdAbhishek:2025dhx}:
\begin{multline}
    \langle J^{\mathrm{a}}\left({\bf k}_1;{\bm \epsilon_1 }\right){\tilde a}^{\mathrm{b}}\left({\bf k}_2;{\bm \epsilon_2 }\right)J^{\mathrm{c}}\left({\bf k}_3;{\bm \epsilon_3 }\right)\rangle = 2 \sin\left(\left(d+2\right)\tfrac{\pi}{2} \right) c^{\text{dS-AdS}}_{\Delta_J}c^{\text{dS-AdS}}_{\Delta_a} c^{\text{dS-AdS}}_{\Delta_J} \\ \times \langle J^{\mathrm{a}}\left({\bf k}_1;{\bm \epsilon_1 }\right){\tilde a}^{\mathrm{b}}\left({\bf k}_2;{\bm \epsilon_2 }\right)J^{\mathrm{c}}\left({\bf k}_3;{\bm \epsilon_3 }\right)\rangle_{\text{EAdS},\,\text{contact}} + O(\mathrm{g}^2).
\end{multline}
 To assemble the Witten diagram we need the gauge boson bulk-to-boundary propagator \eqref{gbbubo}. To avoid clutter we do not display explicitly terms proportional to ${\bm \epsilon}_2\cdot {\bf k}_2$ and ${\bm \epsilon}_1\cdot {\bf k}_1$.\footnote{Due to gauge invariance the helicity 0 component of $\tilde{a}$ is unphysical.} Employing the Mellin-Barnes representation \eqref{BesselKMellin} of the Bessel-$K$ function, one finds the relevant terms in the Mellin amplitude \eqref{MellinA} are
\begin{multline}
{\cal A}^{\text{EAdS},\,\text{contact}}_{\Delta_J \Delta_a \Delta_J}\left(s_1,{\bf k}_1,{\bm \epsilon}_1;s_2,{\bf k}_2,{\bm \epsilon}_2;s_3,{\bf k}_3,{\bm \epsilon}_3\right)=-i\mathrm{g} f^{\mathrm{a}\mathrm{b}\mathrm{c}}\frac{\pi^{\frac{3}{2}}}{\Gamma\left(2-\frac{d}{2}\right)\Gamma \left(\frac{d}{2}\right)^2}\lim_{z_0 \to 0}\left[{\bm \epsilon}_1\cdot\textbf{k}_{23}{\bm \epsilon}_2\cdot{\bm \epsilon}_3\right.\nonumber\\  \left.+{\bm \epsilon}_2\cdot\textbf{k}_{31}{\bm \epsilon}_3\cdot{\bm \epsilon}_1+{\bm \epsilon}_1\cdot{\bm \epsilon}_2\left({\bm \epsilon}_3\cdot\textbf{k}_{12}+\frac{4 (s_1-s_2) {\bm \epsilon}_3\cdot \textbf{k}_3 }{d-6+4 s_3}\right)\right]z^{\frac{d}{2}-2 (s_1+s_2+s_3)+1}_0,
\end{multline}
where the divergence of $J\left({\bf k}_3;{\bm \epsilon_3 }\right)$ gives
\begin{multline}
    \left({\bf k}_3 \cdot \partial_{{\bm \epsilon}_3}\right) {\cal A}^{\text{EAdS},\,\text{contact}}_{\Delta_J \Delta_a \Delta_J}\left(s_1,{\bf k}_1,{\bm \epsilon}_1;s_2,{\bf k}_2,{\bm \epsilon}_2;s_3,{\bf k}_3,{\bm \epsilon}_3\right)\\=i\mathrm{g}f^{\mathrm{a}\mathrm{b}\mathrm{c}}\frac{\pi^{\frac{3}{2}}}{\Gamma\left(2-\frac{d}{2}\right)\Gamma \left(\frac{d}{2}\right)^2}(d-2) {\bm \epsilon}_1\cdot{\bm \epsilon}_2\,\lim_{z_0 \to 0}(s_1-s_2) z_0^{\frac{d}{2}-2 s_1-2 s_2-2 s_3-1}.
\end{multline}
In this case the Mellin integrals \eqref{MellinA} are evaluated by taking the residues of poles at:
\begin{equation}
    s_1 = \pm \frac{1}{2} \left(\frac{d}{2}-1\right) , \qquad  s_2 = \mp \frac{1}{2} \left(\frac{d}{2}-1\right), \qquad  s_3 =\frac{1}{2} \left(\frac{d}{2}-1\right). 
\end{equation}
These give\footnote{The divergence for even $d$ can be dealt with using the prescription given in \cite{MdAbhishek:2025dhx}. On a practical level, the expression for even $d$ is given by the finite part.} 
\begin{multline}\label{anomWYM}
    \left({\bf k}_3 \cdot \partial_{{\bm \epsilon}_3}\right)  \langle J^{\mathrm{a}}\left({\bf k}_1;{\bm \epsilon_1 }\right){\tilde a}^{\mathrm{b}}\left({\bf k}_2;{\bm \epsilon_2 }\right)J^{\mathrm{c}}\left({\bf k}_3;{\bm \epsilon_3 }\right)\rangle=-i\mathrm{g} f^{\mathrm{a}\mathrm{b}\mathrm{c}} \frac{{\bm \epsilon}_1\cdot{\bm \epsilon}_2 \csc ^2\left(\frac{\pi  d}{2}\right)}{16 (d-2)}\,\left[1-{k}_1^{d-2} {k}_2^{2-d}\right]\\+O(\mathrm{g}^2),
\end{multline}
where, as before, we dropped contact terms and components longitudinal to ${\bm \epsilon}_{1,2}$. This signals non-conservation of $J_i$ in the form:
\begin{equation}
    \partial \cdot J^{\mathrm{a}} = \alpha f^{\mathrm{a}\mathrm{b}\mathrm{c}} \left[ {\tilde a}^{\mathrm{b}} J^{\mathrm{c}} \right]_{0,0} + O(\mathrm{g}^2),
\end{equation}
which arises from multiplet recombination with the double-trace operator
\begin{equation}\label{DTYM}
    \left[ {\tilde a}^{\mathrm{b}} J^{\mathrm{c}} \right]_{0,0} = {\tilde a}^{\mathrm{b}} \cdot J^{\mathrm{c}}.
\end{equation}
The anomalous Ward identity \eqref{anomWYM} fixes $\alpha$ to be:
\begin{equation}\label{alphaYM}
    \alpha = -\mathrm{g} \frac{\left(d-2\right)}{4}.
\end{equation}

\subsection{Pure Einstein Gravity}
\label{subsec::EG}

The final example we consider is pure Einstein gravity on dS$_{d+1}$. In the temporal gauge the on-shell vertex for the weak field fluctuations $h_{ij}$ is simply (see e.g. \cite{MdAbhishek:2025dhx} equation (6.3)): 
\begin{equation}\label{cubicGR}
     {\cal V}_{hhh} \approx  \frac{1}{2}\kappa (-\eta)^8\left[ -\frac{1}{2}\delta^{i i_1} \delta^{k k_1} \delta^{l l_1} \delta^{j j_1} + \delta^{i i_1} \delta^{j k_1} \delta^{j_1 k} \delta^{l_1 l}\right]h_{i_1 j_1} \partial_i h_{k_1 l_1} \partial_{j} h_{k l}.
\end{equation}
On the late time boundary we have the operators:
\begin{align}
    h\left(\eta \to 0, {\bf x};{\bm \epsilon }\right) &= \eta^{-2}{\tilde h}\left({\bf x};{\bm \epsilon }\right)+ \left(-\eta\right)^{d-2} T\left({\bf x};{\bm \epsilon }\right),
\end{align}
where ${\tilde h}\left({\bf x};{\bm \epsilon }\right)$ is the boundary graviton and $T\left({\bf x};{\bm \epsilon }\right)$ the boundary stress tensor, which is classically conserved with scaling dimension $\Delta_T = d$.

\vskip 4pt
To study the quantum corrections to $T_{ij}$ we consider its the three-point function with itself and the boundary graviton ${\tilde h}_{ij}$. The contact diagram contribution to the latter, generated by the cubic vertex \eqref{cubicGR}, can be expressed in terms of the corresponding contact Witten diagram in EAdS via \cite{MdAbhishek:2025dhx}:
\begin{multline}
    \langle T\left({\bf k}_1;{\bm \epsilon_1 }\right){\tilde h}\left({\bf k}_2;{\bm \epsilon_2 }\right)T\left({\bf k}_3;{\bm \epsilon_3 }\right)\rangle = 2 \sin\left(\left(d+2\right)\tfrac{\pi}{2} \right) c^{\text{dS-AdS}}_{\Delta_T}c^{\text{dS-AdS}}_{\Delta_h} c^{\text{dS-AdS}}_{\Delta_T} \\ \times \langle T\left({\bf k}_1;{\bm \epsilon_1 }\right){\tilde h}\left({\bf k}_2;{\bm \epsilon_2 }\right)T\left({\bf k}_3;{\bm \epsilon_3 }\right)\rangle_{\text{EAdS},\,\text{contact}} + O(\mathrm{\kappa}^2).
\end{multline}

\vskip 4pt
 Proceeding as in the previous sections, the relevant terms in the Mellin amplitude \eqref{MellinA} are 
\begin{multline}
{\cal A}^{\text{EAdS},\,\text{contact}}_{\Delta_T \Delta_h \Delta_T}\left(s_1,{\bf k}_1,{\bm \epsilon}_1;s_2,{\bf k}_2,{\bm \epsilon}_2;s_3,{\bf k}_3,{\bm \epsilon}_3\right)\\=-\kappa\frac{\pi^{\frac{3}{2}}}{2\Gamma \left(\frac{d}{2}+1\right)^2 \Gamma \left(1-\frac{d}{2}\right)}({\bm \epsilon}_1\cdot\textbf{k}_{23}{\bm \epsilon}_2\cdot{\bm \epsilon}_3+{\bm \epsilon}_2\cdot\textbf{k}_{31}{\bm \epsilon}_3\cdot{\bm \epsilon}_1+{\bm \epsilon}_3\cdot\textbf{k}_{12}{\bm \epsilon}_1\cdot{\bm \epsilon}_2)\,\\ \times \lim_{z_0 \to 0}\left[{\bm \epsilon}_1\cdot\textbf{k}_{23}{\bm \epsilon}_2\cdot{\bm \epsilon}_3+{\bm \epsilon}_2\cdot\textbf{k}_{31}{\bm \epsilon}_3\cdot{\bm \epsilon}_1+{\bm \epsilon}_3\cdot\textbf{k}_{12}{\bm \epsilon}_1\cdot{\bm \epsilon}_2+\frac{4 (s_1-s_2) {\bm \epsilon}_3\cdot \textbf{k}_3 {\bm \epsilon}_1\cdot {\bm \epsilon}_2}{d-6+4 s_3}\right]z^{\frac{d}{2}-2 (s_1+s_2+s_3)+2}_0,
\end{multline}
where as before we do not display explicitly terms proportional to ${\bm \epsilon}_2\cdot {\bf k}_2$ and ${\bm \epsilon}_1\cdot {\bf k}_1$, and also to $\left({\bm \epsilon}_3\cdot \textbf{k}_3\right)^2$ which contributes lower helicity terms. Evaluating the divergence of $T({\bf k}_3,{\bm \epsilon}_3)$, one obtains
\begin{multline}
   \left({\bf k}_3 \cdot D_{{\bm \epsilon_3}}\right) {\cal A}^{\text{EAdS},\,\text{contact}}_{\Delta_T \Delta_h \Delta_T}\left(s_1,{\bf k}_1,{\bm \epsilon}_1;s_2,{\bf k}_2,{\bm \epsilon}_2;s_3,{\bf k}_3,{\bm \epsilon}_3\right)=\kappa d\frac{\pi^{\frac{3}{2}}}{\Gamma \left(\frac{d}{2}+1\right)^2 \Gamma \left(1-\frac{d}{2}\right)} \\
   \times  (2 {\bm \epsilon}_1\cdot \textbf{k}_2 {\bm \epsilon}_2\cdot{\bm \epsilon}_3+{\bm \epsilon}_1\cdot{\bm \epsilon}_2 ({\bm \epsilon}_3\cdot\textbf{k}_1-{\bm \epsilon}_3\cdot\textbf{k}_2)+2 {\bm \epsilon}_1\cdot{\bm \epsilon}_3{\bm \epsilon}_2\cdot \textbf{k}_3){\bm \epsilon}_1\cdot{\bm \epsilon}_2\\ \times \,\lim_{z_0 \to 0}(s_1-s_2) z_0^{\frac{d}{2}-2 s_1-2 s_2-2 s_3}.
\end{multline}
The Mellin integrals \eqref{MellinA} are evaluated from the residues of poles at:
\begin{equation}
    s_1 = \pm \frac{d}{4}, \qquad s_2 = \mp \frac{d}{4}, \qquad s_3 = \frac{d}{4}.
\end{equation}
Up to longitudinal and contact terms this gives\footnote{The divergence for even $d$ can be dealt with using the prescription given in \cite{MdAbhishek:2025dhx}. On a practical level, the expression for even $d$ is given by the finite part.} 
\begin{multline}\label{anomWEH}
   \left({\bf k}_3 \cdot D_{{\bm \epsilon}_3}\right) \langle T\left({\bf k}_1;{\bm \epsilon_1 }\right){\tilde h}\left({\bf k}_2;{\bm \epsilon_2 }\right)T\left({\bf k}_3;{\bm \epsilon_3 }\right)\rangle= -\kappa\frac{\csc^2\left(\frac{\pi  d}{2}\right)}{16 d}\\ \times (2 {\bm \epsilon}_1\cdot\textbf{k}_2 {\bm \epsilon}_2\cdot{\bm \epsilon}_3+{\bm \epsilon}_1\cdot{\bm \epsilon}_2 ({\bm \epsilon}_3\cdot\textbf{k}_1-{\bm \epsilon}_3\cdot\textbf{k}_2)+2 {\bm \epsilon}_1\cdot{\bm \epsilon}_3{\bm \epsilon}_2\cdot\textbf{k}_3)k_1^d k_2^{-d}.
\end{multline}
This signals non-conservation of $T_{ij}$ in the form: 
\begin{align}
    \partial^{i} T_{ij} =  \alpha [ T {\tilde h}]_{0,j}+O(\kappa^2)\,,
\end{align}
arising from multiplet recombination with the spin-1 double-trace operator: 
\begin{multline}
    [T{\tilde h}]_{0,1}\left({\bf k}_1+{\bf k}_2;{\bm \epsilon}_3\right) = i{\bm \epsilon}_3\cdot {\bf k}_{12}  T_{ij}\left({\bf k}_1\right){\tilde h}^{ij}\left({\bf k}_2\right) \\ + 2i\left(k_1+k_2\right)^i \epsilon^j_3\left(  T^{l}{}_{j}\left({\bf k}_1\right){\tilde h}_{il}\left({\bf k}_2\right) - T^{l}{}_{i}\left({\bf k}_1\right){\tilde h}_{jl}\left({\bf k}_2\right) \right),
\end{multline}
which we give up to longitudinal terms. Note that the second line, being proportional to the momentum, is a descendant. The anomalous Ward identity \eqref{anomWEH} fixes $\alpha$ to be: 
\begin{equation}\label{alphaEH}
    \alpha = -\kappa \frac{d}{4} .
\end{equation}

\section{Anomalous dimensions}
\label{subsec::ad}

We have seen that currents on the dS boundary tend to acquire anomalous dimensions at the quantum level via multiplet recombination with double-trace operators of the form:\begin{subequations}\label{recomb}
 \begin{align}
    \partial^i J_i &= \alpha_{J} \left[{\cal O}_{\Delta_+} {\cal O}_{\Delta_-}\right]_{0} + O(g^2),\\
     \partial^i T_{ij} & = \alpha_{T} \left[{\cal O}_{\Delta_+} {\cal O}_{\Delta_-}\right]_{0,j} + O(g^2),
\end{align}   
\end{subequations}
where in the previous section $\alpha_{J}$ was extracted for scalar QED and Yang-Mills; and $\alpha_T$ for a minimally coupled scalar and pure Einstein gravity.

\vskip 4pt
The corresponding anomalous dimensions $\gamma_J$ and $\gamma_T$ can then be read off from the two-point functions of $\partial^i J_i$ and $\partial^i T_{ij}$, where the highest helicity components are given by the expressions:\footnote{Note that the normalisation is chosen consistently with our definition \eqref{ThomasD} of the Thomas-D operator.}
\begin{subequations}\label{2ptanom}
 \begin{align}
    \left\langle \partial\cdot J({\bf k}) \partial\cdot J(-{\bf k})\right\rangle&=\gamma_J\,\frac{   \Gamma \left(1-\frac{d}{2}\right) \Gamma \left(2-\frac{d}{2}\right)}{2\pi } \left(\frac{k}2\right)^d,\\
    \left\langle \partial\cdot T({\bf k};{\bm \epsilon}_1)\partial\cdot T(-{\bf k};{\bm \epsilon}_2)\right\rangle&= \gamma_T\frac{\Gamma \left(1-\frac{d}{2}\right) \Gamma \left(-\frac{d}{2}\right) }{\pi}\,{\bm \epsilon}_1\cdot {\bm \epsilon}_2\,\left(\frac{k}2\right)^{d+2}\,.
\end{align}   
\end{subequations}
These can be extracted from the expression \cite{Sleight:2019hfp} for the late-time free theory two-point function of a spin-$s$ field with scaling dimension $\Delta = \Delta_{J/T}+\gamma_{J/T}$:
\begin{multline}
    \left\langle {\cal O}_{\Delta,s}({\bf k};{\bm \epsilon}_1){\cal O}_{\Delta,s}(-{\bf k};{\bm \epsilon}_2)\right\rangle=\pi ^{d/2}\sum^s_{n=0}\binom{s}{n} \left(\frac{k}{2}\right)^{2 \Delta -d}\frac{(\Delta -1)_n \Gamma \left(\frac{d}{2}+s-n-\Delta \right)}{\Gamma (s+\Delta )}\\ \times \left(\frac{2 {\bm \epsilon}_1\cdot {\bf k} \,{\bm \epsilon}_2\cdot {\bf k}}{k^2}\right)^{s-n}({\bm \epsilon}_1\cdot {\bm \epsilon}_2)^n\,.
\end{multline}
Due to the multiplet recombination \eqref{recomb}, these two-point functions can be expressed in terms of two-point functions of the corresponding double-trace operators,
\begin{align}
    \left\langle \partial\cdot J \partial\cdot J\right\rangle&=\alpha^2_J\,\left\langle \left[{\cal O}_{\Delta_+} {\cal O}_{\Delta_-}\right]_{0} \left[{\cal O}_{\Delta_+} {\cal O}_{\Delta_-}\right]_{0}\right\rangle, \\
    \left\langle \partial^{i}T_{i i_1}\,\partial^{j}T_{j i_2}\right\rangle&= \alpha^2_T\,\left\langle \left[{\cal O}_{\Delta_+} {\cal O}_{\Delta_-}\right]_{0,i_1} \left[{\cal O}_{\Delta_+} {\cal O}_{\Delta_-}\right]_{0,i_2}\right\rangle.
\end{align}
The latter can be extracted in each case by taking appropriate derivatives in $\lambda_i$ and ${\bar \lambda}_i$ of the generating function: 
\begin{multline}\label{genformDT}
       \int \frac{{\rm d}^d {\bf k}_1}{\left(2\pi \right)^d} \frac{{\rm d}^d {\bf k}_2}{\left(2\pi \right)^d}\, e^{i\lambda_1 {\bm \epsilon}_3\cdot {\bf k}_1+i\lambda_2 {\bm \epsilon}_3\cdot {\bf k}_2-i\bar{\lambda}_1 \bar{{\bm \epsilon}}_3\cdot {\bf k}_1-i{\bar \lambda}_2 \bar{{\bm \epsilon}}_3\cdot {\bf k}_2}\,e^{i({\bf k}_1+{\bf k}_2)\cdot {\bf x}_{12}}\\ \times \langle {\cal O}_{\Delta_1}\left({\bf k}_1\right){\cal O}_{\Delta_1}\left(-{\bf k}_1\right) \rangle\langle {\cal O}_{\Delta_2}\left({\bf k}_2\right){\cal O}_{\Delta_2}\left(-{\bf k}_2\right) \rangle=\int \frac{{\rm d}^d {\bf k}}{\left(2\pi \right)^d}\, A(\lambda_i,\bar{\lambda}_i|{\bf k})e^{i{\bf k}\cdot {\bf x}_{12}},
\end{multline}
where performing the change of variable ${\bf k}_1\to {\bf k}_1-{\bf k}_2$ to evaluate the integral over ${\bf k}_2$ gives
\begin{align}
    A(\lambda_i,\bar{\lambda}_i|{\bf k})=&16\pi^{2-\frac{d}{2}}\Gamma \left(\tfrac{d}{2}-\Delta_1\right) \Gamma \left(\tfrac{d}{2}-\Delta_2\right)\\
    & \times \sum^\infty_{n=0}\sum^\infty_{m=0}\,\frac{ 2^{m} \Gamma (m+\Delta_1) \Gamma (m+n+\Delta_2) \Gamma \left(\frac{d-2 (m+\Delta_1+\Delta_2)}{2}\right)}{m! n!  \Gamma (2 m+n+\Delta_1+\Delta_2)}\nonumber\\&\times e^{i (\lambda_1 {\bf k}\cdot {\bm \epsilon}_3-{\bar \lambda}_1 {\bf k}\cdot {\bm {\bar \epsilon}}_3)}((\lambda_2-\lambda_1) ({\bar \lambda}_2-\lambda_1) {\bm \epsilon}_3\cdot {\bar {\bm \epsilon}}_3)^m\nonumber \\
    &\times(i ((\lambda_2-\lambda_1) {\bf k}\cdot {\bm \epsilon}_3-({\bar \lambda}_2-{\bar \lambda}_1) {\bf k}\cdot {\bar {\bm \epsilon}}_3))^n \,\left(\frac{k}{2}\right)^{2 (\Delta_1+\Delta_2+m)-d}.
\end{align}
Restricting to the highest helicity components (setting ${\bf k}\cdot {\bm \epsilon}_3 = {\bf k}\cdot {\bar {\bm \epsilon}}_3$=0), one can read off expressions for the corresponding anomalous dimensions $\gamma_J$ and $\gamma_T$ by comparing with \eqref{2ptanom}, which we give in the following. To the best of our knowledge, these expressions are new. They are related to their EAdS counterparts via \eqref{dSEAdSanom} and give the one-loop correction to the mass of the corresponding gauge boson/graviton via \eqref{massform}.

\subsection{Scalar QED}
Let us consider scalar QED first. As shown explicitly in section \ref{subsec::scalarqed}, the boundary current $J_i$ exhibits the following mixing:
\begin{equation}
     \partial\cdot J=\alpha_J [\mathcal{O}_{\Delta_+}\mathcal{O}^\dagger_{\Delta_-}]_{0}+\beta_J [\mathcal{O}^\dagger_{\Delta_+}\mathcal{O}_{\Delta_-}]_{0}+O(e^2),
\end{equation}
where $\alpha_J$ and $\beta_J$ were given in \eqref{alphaJqed}. This results in non-conservation of the two-point function in the form,
\begin{align}
    \langle \partial \cdot J \partial \cdot J \rangle = 2\alpha_J\beta_J \langle {\cal O}_{\Delta_+}{\cal O}^\dagger_{\Delta_+} \rangle  \langle {\cal O}_{\Delta_-}{\cal O}^\dagger_{\Delta_-} \rangle + O(e^3).
\end{align}
Going to momentum space and comparing with \eqref{2ptanom}, the corresponding anomalous dimension $\gamma_J$ can be read off to be:
\begin{equation}
   \gamma_J = e^2 \frac{\Gamma (1-i \mu ) \Gamma (1+i \mu ) \Gamma \left(\frac{d}{2}-i \mu \right) \Gamma \left(\frac{d}{2}+i \mu \right)}{16\pi ^{\frac{d}{2}+1}\Gamma \left(2-\frac{d}{2}\right) \Gamma (d+1)}.
\end{equation}
We see that for $\mu \in \mathbb{R}$, which corresponds to a massive scalar field in dS, the anomalous dimension $\gamma_J$ is positive/negative when $c^{\text{dS-AdS}}_{\Delta_J}$ is positive/negative. The latter is the kinetic term of the corresponding gauge field in EAdS \cite{MdAbhishek:2025dhx}.

\subsection{Scalar field minimally coupled to gravity}

For a scalar field minimally coupled to gravity, from section \ref{subsec::scmingrav} we have
\begin{align}
    \partial^{i} T_{ij} =  \alpha_T \left[{\cal O}_{\Delta_+} {\cal O}_{\Delta_-}\right]_{0,j}+O(\kappa^2)\,,
\end{align}
where $\alpha_T$ is given in \eqref{scgravalpha} and the double-trace operator in \eqref{DTscgrav}. Following the same steps and extracting the double-trace two-point function from the generating formula \eqref{genformDT}, the expression for the anomalous dimension of the stress tensor is:
\begin{align}
    \gamma_T=\frac{\kappa ^2 \Gamma (1-i \mu ) \Gamma (1+i \mu ) \Gamma \left(\frac{d}{2}-i \mu +1\right) \Gamma \left(\frac{d}{2}+i \mu +1\right)}{4 \pi ^{\frac{d}{2}+1} \Gamma \left(1-\frac{d}{2}\right) \Gamma (d+3)}.
\end{align}
As in scalar QED, for a massive scalar field $\mu \in \mathbb{R}$ the anomalous dimension $\gamma_T$ is positive/negative when $c^{\text{dS-AdS}}_{\Delta_T}$ is positive/negative. The latter is the kinetic term of the corresponding graviton in EAdS \cite{MdAbhishek:2025dhx}.

\subsection{Yang-Mills theory}

For Yang-Mills theory, from section \ref{subsec::YM} we have
\begin{equation}
   \partial \cdot J^{\mathrm{a}} = \alpha_J f^{\mathrm{a}\mathrm{b}\mathrm{c}} \left[ {\tilde a}^{\mathrm{b}} J^{\mathrm{c}} \right]_{0,0} + O(\mathrm{g}^2),
\end{equation}
where $\alpha_J$ is given in \eqref{alphaYM} and the double-trace operator in \eqref{DTYM}. In this case the corresponding anomalous dimension is:
\begin{align}
    \gamma_J=\mathrm{g}^2\frac{\left(d^2-5 d+5\right)  \Gamma \left(\frac{d}{2}-1\right)}{32 \pi ^{\frac{d}{2}+1} (d-2) (d-1)}.
\end{align}

\subsection{Pure Einstein Gravity}

For Pure Einstein Gravity, from section \ref{subsec::EG} we have
\begin{align}
    \partial^{i} T_{ij} =  \alpha_T [ T {\tilde h}]_{0,j}+O(\kappa^2)\,,
\end{align}
where $\alpha_T$ is given in \eqref{alphaEH} and for convenience we repeat the form of the spin-1 double-trace operator below (modulo longitudinal terms) : 
\begin{multline}
    [T{\tilde h}]_{0,1}\left({\bf k}_1+{\bf k}_2;{\bm \epsilon}_3\right) = i{\bm \epsilon}_3\cdot {\bf k}_{12}  T_{ij}\left({\bf k}_1\right){\tilde h}^{ij}\left({\bf k}_2\right) \\ + 2i\left(k_1+k_2\right)^i \epsilon^j_3\left(  T^{l}{}_{j}\left({\bf k}_1\right){\tilde h}_{il}\left({\bf k}_2\right) - T^{l}{}_{i}\left({\bf k}_1\right){\tilde h}_{jl}\left({\bf k}_2\right) \right).
\end{multline}
This case presents an additional subtlety since the two-point function of the above double-trace operator is divergent, signaling an IR divergence that requires renormalisation. This can be resolved by the fact that the boundary graviton ${\tilde h}$ has vanishing scaling dimension $\Delta_h=0$, such that composites formed from a single $T_{ij}$ with ${\tilde h}_{ij}$ also have dimension $\Delta_T$. The boundary stress tensor can therefore be redefined in the following way:
\begin{multline}
    \left({\bf k}_1+{\bf k}_2\right)\cdot D_{{\bm \epsilon}_3} \left[\underbrace{T({\bf k}_1+{\bf k}_2;{\bm \epsilon}_3)+\alpha_T\, C\, {\bm \epsilon}_2\cdot {\bm \epsilon}_3 {\bm \epsilon}_1\cdot {\bm \epsilon}_3 \,{\bm \epsilon}_1 \cdot {\bm \epsilon}_2}_{\bar{T} }\right]\\=\alpha_T\,{\bm \epsilon}_1\cdot {\bm \epsilon}_2\left[{\bm \epsilon}_3\cdot {\bf k}_{12} \,{\bm \epsilon}_1\cdot {\bm \epsilon}_2 \right. \\ \left.+ (2-C){\bm \epsilon}_3\cdot {\bm \epsilon}_1 {\bm \epsilon}_2\cdot ({\bf k}_1+{\bf k}_2)-(2+C){\bm \epsilon}_3\cdot {\bm \epsilon}_2 {\bm \epsilon}_1\cdot ({\bf k}_1+{\bf k}_2)\right]\,,
\end{multline}
where $C$ is an arbitrary constant.
This equation can be interpreted schematically as follows:
\begin{align}
    \partial \cdot (T+hT)=[hT]
\end{align}
so that at order $\kappa^2$ the stress tensor acquires both an anomalous dimension and a non-linear deformation. The redefinition is fixed by requiring that the double-trace operators on the right-hand side of the equation have a finite two-point function which, using \eqref{genformDT}, sets $C=2$. It would be interesting to investigate the regularisation of the IR divergence more systematically.

\vskip 4pt
Proceeding as before, the anomalous dimension of the stress tensor is given by:
\begin{align}
    \gamma_T=\kappa ^2\frac{2^{d-2} (d-2) d^3 (d+1) (d (3 (d-3) d-10)+8) \pi ^{\frac{d}{2}+1} }{(d-1)^2 (d+2) \Gamma \left(1-\frac{d}{2}\right)^2 \Gamma \left(\frac{d}{2}\right)}.
\end{align}

\newpage 
\section{Discussion and conclusion}
\label{sec::discussion}

In this work we have shown that, unlike in AdS, no exact global symmetry survives at the late-time boundary of de Sitter space. Boundary radiation fluctuations induce multiplet recombination, giving anomalous dimensions to boundary currents and the stress tensor. We verified this mechanism explicitly in scalar QED, Einstein gravity, and Yang–Mills theory, and argued that it extends to all gauge fields, including higher-spin and partially massless cases. From the perspective of the EAdS reformulation \cite{Sleight:2020obc,Sleight:2021plv} of dS late-time correlators, the picture that emerges is a Higgs-like mechanism in which Dirichlet modes acquire small masses through mixing with composite operators formed from a pair of shadow operators, while Neumann modes remain protected by gauge symmetry.

\vskip 4pt
We conclude by mentioning a couple of conceptual subtleties.

\paragraph{Gauge choice and locality.} Our analysis was carried out in temporal gauge, where the field algebra is local. This makes it possible to work with local operator algebras, which have proved powerful in QFT—for instance, in establishing analyticity properties of vacuum correlation functions and renormalisation of QFT. The drawback is that local gauges violate Gauss’ law. In QED, imposing Gauss’ law to find local physical states projects to the zero-charge superselection sector. This sector is still non-trivial in gauge theories, but in gravity every state carries energy, raising sharper issues related to the absence of local observables altogether.

\vskip 4pt
These difficulties are standard in gauge theories (see e.g.~\cite{strocchiBook}): they can only be addressed either by abandoning strict locality in favor of nonlocal gauges, or by dressing local fields in a Hilbert–Krein topology so that charged states appear in the closure of the local algebra. In this setting one assumes that the closure of the Hilbert space includes charged fields. Haag and Kastler argued long ago \cite{Haag:1963dh} that, at least in QED, the zero-charge local sector already contains the full physical content of the theory. Extending such reasoning to gravity is subtler, since every state has nonzero energy and the necessary nonlocal dressings extend to infinity. Still, one can attempt to equip the local field algebra with an appropriate topology so that physical states can be approximated arbitrarily well.

\paragraph{Gauge invariance.} For the reasons discussed above, the three-point functions evaluated in this work are not gauge-invariant observables: they are invariant only under linearised gauge transformations. Full gauge invariance requires dressing charged local operators with nonlocal factors—such as Wilson lines in QED and QCD,
\begin{align}
    &\phi^*(x)\phi(y)\to \phi^*(x)e^{i\int_x^y A_\mu dx^\mu}\phi(y)\,,& &\phi(x)\to e^{i\int_\infty^x A_\mu dx^\mu}\phi(x)\,,
\end{align}
with analogous (and more intricate) dressings in gravity \cite{Giddings:2019wmj}. Nevertheless, the local operator algebra, suitably completed, has been argued to contain sufficient information to reconstruct the Hilbert space of the theory \cite{strocchiBook}. More recently, \cite{Grassi:2024vkb} proposed an automorphism relating the local algebra to the nonlocal charged algebras of different superselection sectors. Such constructions are central in gravity, where they connect to the definition of subregion/island algebras and entropy calculations. Once operators are dressed and hence nonlocal, it is in fact no longer guaranteed that the algebra associated to a compact subregion commutes with operators supported in the complement, since the dressing necessarily extends to infinity and overlaps with the outside region. This tension has important implications: if subregion algebras fail to commute with their complements, entropy estimates and degrees-of-freedom counting break down. In \cite{Grassi:2024vkb} it was argued that the non-local algebra is isomorphic to the local one, at least at a formal level, allowing to justify more clearly these statements.

\paragraph{No Higgs mechanism for gravity in dS?} A standard argument against the Higgs mechanism for gravity in de Sitter space invokes the Higuchi bound \cite{Higuchi:1986py}, which states that unitary representations require the graviton mass to lie above the partially massless spin-2 point. Below this threshold the representation becomes non-unitary, with the strictly massless case appearing as a discrete limit. This seems to forbid the graviton from acquiring a small mass. This argument, however, overlooks the field doubling inherent to the in-in formalism. With this doubling one has a field whose late-time limit encodes the boundary graviton, which remains protected by gauge invariance, and another field giving the the boundary stress tensor. The latter can become unstable due to cosmic expansion, effectively acquiring a complex mass and turning into a resonance on the second sheet. In this way the Dirichlet component of the graviton is destabilised without contradicting the Higuchi bound, while the Neumann component—the true boundary graviton—remains safeguarded by gauge symmetry.

\section*{Acknowledgments}

This research was supported by the European Union (ERC grant ``HoloBoot'', project number 101125112),\footnote{Views and opinions expressed are however those of the author(s) only and do not necessarily reflect those of the European Union or the European Research Council. Neither the European Union nor the granting authority can be held responsible for them.} by the MUR-PRIN grant No. PRIN2022BP52A (European Union - Next Generation EU) and by the INFN initiative STEFI.

\newpage

\appendix

\section{Boundary correlators in Mellin space}
\label{app::Mellinspace}

In this appendix we review the relevant aspects of (EA)dS boundary correlators in Mellin space---for further details see \cite{Sleight:2021plv} and references therein.

\vskip 4pt
Consider the boundary three-point functions of operators ${\cal O}_{\Delta_i}\left({\bf x}_i;{\bm \epsilon}_i\right)$ with scaling dimension $\Delta_i$. Combining Fourier space \eqref{FT} and Mellin space \eqref{MT}, this can be expressed in the form
\begin{multline}
  \langle {\cal O}_{\Delta_1}\left({\bf k}_1;{\bm \epsilon}_1\right){\cal O}_{\Delta_2}\left({\bf k}_2;{\bm \epsilon}_2\right) {\cal O}_{\Delta_3}\left({\bf k}_3;{\bm \epsilon}_3\right) \rangle\\= \left(2\pi\right)^d \delta^{(d)}\left({\bf k}_1+{\bf k}_2+{\bf k}_3\right)  \langle {\cal O}_{\Delta_1}\left({\bf k}_1;{\bm \epsilon}_1\right){\cal O}_{\Delta_2}\left({\bf k}_2;{\bm \epsilon}_2\right) {\cal O}_{\Delta_3}\left({\bf k}_3;{\bm \epsilon}_3\right) \rangle^\prime , 
\end{multline}
\begin{multline}\label{MBrep3pt}
    \langle {\cal O}_{\Delta_1}\left({\bf k}_1;{\bm \epsilon}_1\right){\cal O}_{\Delta_2}\left({\bf k}_2;{\bm \epsilon}_2\right) {\cal O}_{\Delta_3}\left({\bf k}_3;{\bm \epsilon}_3\right) \rangle^\prime =  \int_{-i\infty}^{+i\infty} \prod^3_{i=1} \frac{{\rm d}s_i}{2\pi i}\,  \frac{1}{2\sqrt{\pi}}\Gamma\left(s_i\pm \tfrac{1}{2}\left(\Delta_i-\tfrac{d}{2}\right)\right)\left(\frac{k_i}{2}\right)^{-2s_i + \Delta_i -\frac{d}{2}}\\ \times {\cal A}_{\Delta_1 \Delta_2 \Delta_3}\left(s_i,{\bf k}_i;{\bm \epsilon}_i\right),
\end{multline}
with Mellin amplitude \cite{Sleight:2021iix}:  
\begin{equation}\label{MellinA}
    {\cal A}_{\Delta_1 \Delta_2 \Delta_3}\left(s_i,{\bf k}_i;{\bm \epsilon}_i\right) = 2 \pi i\, \delta\left(\tfrac{x}{2}-2s_1 -2s_2-2s_3\right) \mathfrak{C}\left(s_i, {\bm \epsilon}_i \cdot {\bf k}_j, {\bm \epsilon}_i \cdot {\bm \epsilon}_j \right).
\end{equation}
The function $\mathfrak{C}\left(s_i, {\bm \epsilon}_i \cdot {\bf k}_j, {\bm \epsilon}_i \cdot {\bm \epsilon}_j \right)$ encodes the tensorial structure and is a rational function of the Mellin variables $s_i$. The linear constraint on the latter is implied by the Dilatation Ward identity, where $x$ encodes the structure of the cubic vertex. Analogous to momentum conservation this arises from the integral over the bulk direction:
\begin{align}
    2 \pi i\, \delta\left(\tfrac{x}{2}-2s_1 -2s_2-2s_3\right) &= \lim_{z_0\to 0} \left[\int^\infty_{z_0} \frac{{\rm d}z}{z} z^{\tfrac{x}{2}-2s_1 -2s_2-2s_3}\right],\\
    &= \lim_{z_0\to 0} \left[-\frac{z_0^{\tfrac{x}{2}-2s_1 -2s_2-2s_3}}{\tfrac{x}{2}-2s_1 -2s_2-2s_3} \right].
\end{align}

We will often make use of the Mellin representation of the modified Bessel function of the second kind: 
\begin{align}\label{BesselKMellin}
    K_{i\nu}(kz):=\int_{-i\infty}^{
    +i\infty}\frac{{\rm d}s}{2\pi i}\left(\frac{zk}{2}\right)^{-2s}\frac{\Gamma(s+\frac{i\nu}{2})\Gamma(s-\frac{i\nu}{2})}{2}.
\end{align}
For example, consider the contact Witten diagram generated by the following non-derivative vertex of scalar fields $\phi_i$ of mass $m^2_i=\Delta_i\left(\Delta_i-d\right)$:
\begin{equation}\label{v12n}
    {\cal V}_{123} = g \phi_1 \phi_2 \phi_3.
\end{equation}
In Fourier space \eqref{FT}, the bulk-to-boundary propagator for a scalar field of mass $m^2=\Delta\left(\Delta-d\right)$ in EAdS$_{d+1}$ takes the form
\begin{equation}
    K^{\text{AdS}}_{\Delta}\left(z;{\bf k}\right)= \left(\frac{k}{2}\right)^{\Delta-\frac{d}{2}}\frac{z^{\frac{d}{2}}}{\Gamma(\Delta-\frac{d}{2}+1)}K_{\Delta-\frac{d}{2}}(kz).
\end{equation}
Using the Mellin representation \eqref{BesselKMellin} of the Bessel-$K$ function, the contact Witten diagram generated by the vertex \eqref{v12n} reads 
\begin{multline}
\langle {\cal O}_{\Delta_1}\left({\bf k}_1\right){\cal O}_{\Delta_2}\left({\bf k}_2\right) {\cal O}_{\Delta_3}\left({\bf k}_3\right) \rangle^\prime =  \int_{-i\infty}^{+i\infty} \prod^3_{i=1} \frac{{\rm d}s_i}{2\pi i}\, \frac{1}{2\sqrt{\pi}} \Gamma\left(s_i\pm \tfrac{1}{2}\left(\Delta_i-\tfrac{d}{2}\right)\right)\left(\frac{k_i}{2}\right)^{-2s_i + \Delta_i -\frac{d}{2}}\\ \times {\cal A}_{\Delta_1 \Delta_2 \Delta_3}\left(s_i,{\bf k}_i\right),
\end{multline}
with Mellin amplitude
\begin{align}
 {\cal A}_{\Delta_1 \Delta_2 \Delta_3}\left(s_i,{\bf k}_i\right)&=-g\left(\prod^3_{i=1}\,\frac{\sqrt{\pi}}{\Gamma\left(\Delta_i-\tfrac{d}{2}+1\right)}\right)\int^\infty_0 \frac{{\rm d}z}{z^{d+1}}\,z^{-\sum\limits^3_{i=1}\left(2s_i-\tfrac{d}{2}\right)},\\
 &=-g \left(\prod^3_{i=1}\,\frac{\sqrt{\pi}}{\Gamma\left(\Delta_i-\tfrac{d}{2}+1\right)}\right) \lim_{z_0\to 0} \left[-\frac{z_0^{\tfrac{x}{2}-2s_1 -2s_2-2s_3}}{\tfrac{x}{2}-2s_1 -2s_2-2s_3} \right],\\
 &= -g \left(\prod^3_{i=1}\,\frac{\sqrt{\pi}}{\Gamma\left(\Delta_i-\tfrac{d}{2}+1\right)}\right)2\pi i\, \delta\left(d+\sum\limits^3_{i=1}\left(2s_i-\tfrac{d}{2}\right)\right).
\end{align}

\section{Partially-massless gauge fields}
\label{subsec::PM}

Theories of partially massless fields in de Sitter space were first discovered for lower spin (spin-2 and 3/2) fields in \cite{Deser:1983mm,Higuchi:1986py,Deser:2000de}, and their higher spin generalisation in \cite{Deser:2001pe,Deser:2001us}. 

\vskip 4pt
Consider a partially massless field of spin-$s$ and depth-$r$, where $r \in \left\{0,\ldots,s-1\right\}$. In the free theory, the corresponding boundary currents $J^{(r)}_{i_1 \ldots i_s}({\bf x})$ have scaling dimension:
\begin{equation}\label{pmcurrentdim}
    \Delta_{J^{(r)}_{s}} = (s-r)+d-2, 
\end{equation}
and
satisfy the partial conservation condition \cite{Dolan:2001ih,Deser:2003gw}:
\begin{equation}
    \partial^{i_s} \ldots \partial^{i_{s-r}} J^{(r)}_{i_1 \ldots i_s} = 0.
\end{equation}
 The corresponding boundary partially massless gauge fields ${\tilde \varphi}^{(r)}_{i_1 \ldots i_s}$ have scaling dimension:
\begin{equation}\label{bdrypm}
    \Delta_{{\tilde \varphi}^{(r)}_{s}} = 2-(s-r).
\end{equation}
The value $r=0$ corresponds to an ordinary massless spin-$s$ gauge field \eqref{mlspinsc}.

\vskip 4pt
The cubic interactions of partially massless totally symmetric fields in (A)dS$_{d+1}$ were classified in \cite{Joung:2011ww,Joung:2012hz}. For non-trivial\footnote{I.e. interactions that are not trivially gauge invariant and therefore induce a global symmetry algebra.} cubic interactions of partially massless fields, we have the following necessary condition on the scaling dimensions $\Delta_i$ and spins $s_i$:
\begin{equation}
    (d-2)+\left(s_i+s_{i+1}-s_{i-1}\right)+\left(\Delta_{i-1}-\Delta_i-\Delta_{i+1}\right)= 2 \mathbb{Z},
\end{equation}
where $\left[i \simeq i+3\right]$. For example, for the three-point function of the current $J^{(r)}_{i_1 \ldots i_s}$ with two scalar operators ${\cal O}_{\Delta_{1,2}}$, the dimensions $\Delta_{1,2}$ are constrained by:
\begin{equation}\label{conspm}
    r+\Delta_1-\Delta_2 = 2 \mathbb{Z},
\end{equation}
and also (see \cite{Joung:2012hz})
\begin{equation}\label{addcont}
    |\Delta_1-\Delta_2| \leq r.
\end{equation}
In this case, there is a potential mixing at the quantum level between the divergence of the current \eqref{pmcurrentdim} and composite double-trace operators \eqref{DT} formed from ${\cal O}_{\Delta_1}$ and ${\cal O}_{d-\Delta_2}$, and ${\cal O}_{\Delta_2}$ and ${\cal O}_{d-\Delta_1}$:
\begin{multline}
    \partial^{i_s} \ldots \partial^{i_{s-r}} J^{(r)}_{i_1 \ldots i_s}  \sim g \left[{\cal O}_{\Delta_1}{\cal O}_{d-\Delta_2}\right]_{m+r,i_1 \ldots i_{s-r-1}}+g \left[{\cal O}_{\Delta_2}{\cal O}_{d-\Delta_1}\right]_{m+r,i_1 \ldots i_{s-r-1}}\\+ O(g^2),
\end{multline}
where $m \in \mathbb{Z}$ parameterises the cubic interaction \eqref{conspm} and the constraint \eqref{addcont} requires $0 \leq m \leq r$. The scaling dimension \eqref{bdrypm} of the boundary partially-massless field is instead protected by the partial gauge symmetry.

\vskip 4pt
A similar discussion can be made for other non-trivial cubic interactions of partially massless fields in the classification \cite{Joung:2011ww,Joung:2012hz}.

\bibliographystyle{JHEP}
\bibliography{refs}

\end{document}